\documentclass[10pt,conference]{IEEEtran}

\IEEEoverridecommandlockouts
\IEEEpubid{\makebox[\columnwidth]{Accepted to Workshop on Technology and Consumer
Protection 2020 (ConPro).\hfill} \hspace{\columnsep}\makebox[\columnwidth]{}}

\usepackage[hyphens]{url}

\usepackage{color}
\usepackage[linesnumbered,ruled]{algorithm2e}
\usepackage{cite}

\usepackage{hyperref}
\usepackage{graphicx}
\usepackage{amsmath}

\def\Snospace~{\S{}}

\begin{document}

\title{On the Data Fight Between Cities and Mobility Providers}

\author{\IEEEauthorblockN{Guillermo Baltra\IEEEauthorrefmark{1},
Basileal Imana\IEEEauthorrefmark{2}, Wuxuan Jiang\IEEEauthorrefmark{3} and
Aleksandra Korolova\IEEEauthorrefmark{4}}
\IEEEauthorblockA{University of Southern California\\
Email: \IEEEauthorrefmark{1}baltra@usc.edu,
\IEEEauthorrefmark{2}imana@usc.edu,
\IEEEauthorrefmark{3}wuxuanji@usc.edu,
\IEEEauthorrefmark{4}korolova@usc.edu}}

\IEEEtitleabstractindextext{
    \begin{abstract}
E-Scooters are changing transportation habits.  In an attempt to oversee scooter
usage, the Los Angeles Department of Transportation has put forth a
specification that requests detailed data on scooter usage from scooter companies.  In this work, we first argue that
L.A.'s data request for using a new specification is not warranted as proposed use cases can be
 met by already existing specifications. Second, we
show that even the existing specification, that requires companies to publish
real-time data of parked scooters, puts the privacy of individuals using the scooters at risk.  We then
propose an algorithm that enables formal privacy and utility guarantees when
publishing parked scooters data, allowing city authorities to meet their use cases while
preserving riders' privacy.
    \end{abstract}
}

\maketitle
\IEEEpubidadjcol
\IEEEdisplaynontitleabstractindextext

\section{Introduction}
E-scooters are changing transportation habits. At the same time, their increased usage 
is bringing new challenges such as clogged side-walks and danger of injury to both riders and pedestrians. 
In an effort to oversee and regulate
scooter companies and scooter use, Los Angeles proposed the Mobility Data
Specification (MDS)~\cite{mds_specs}, which requires the scooter companies to
provide real-time data of scooter locations and a historical view of all scooter
operations.

The requirement to provide data according to MDS has not been positively
received by dockless scooter companies
and privacy advocates. The scooter companies argue that such data gives city governments 
an ability to track individual trips, which may have negative privacy implications for individuals~\cite{slate}.
Advocacy groups, such as the Electronic Frontier Foundation and the
Center for Democracy and Technology, have also expressed concerns about rider
privacy~\cite{eff}\cite{cdt} and requested information on how
the Los Angeles Department of Transportation (LADOT) intends to aggregate,
de-identify, and anonymize MDS data if it is to be made public. 
Overall, the question of the data that is necessary for the scooter companies to provide for the cities to oversee scooter usage, 
while at the same time protecting privacy of riders, is highly contested.

Several cities in the US have implemented ad-hoc mechanisms to anonymize scooter
data when sharing it publicly.  For instance, Louisville, KY~\cite{louisville}
aggregates GPS locations
at three decimals for latitude and longitude, which equates to aggregation at approximately a block
level. Austin, TX~\cite{austin} replaces GPS coordinates with census tract IDs,
and Minneapolis, MN~\cite{minneapolis} follows a similar approach but replaces
GPS coordinates with street centerline IDs. Yet, these approaches fail to
protect privacy~\cite{culnane2019stop}. 

\textbf{Contributions: }
In this work we make three contributions to the debate around the scooter data sharing.
\begin{itemize}
\item First, \textit{we argue that Los Angeles's (LA) proposal for the new MDS
specification is not
warranted} (\autoref{sec:you_have_what_you_need}). We do this by showing that LA's proposed data use cases can be met using data that is already
being published as part of an existing specification called GBFS (which was proposed in 2015 and is
already implemented by most scooter companies). We focus on two of LA's main use cases:
determining the number of scooters operating in a region and determining the distribution of scooters across
neighborhoods.
\item Second, \textit{we show that the currently publicly available GBFS data puts individual rider privacy at risk} (\autoref{sec:attacks}).
We demonstrate that this data is susceptible to a trip reconstruction attack, contrary to the claim
by GBFS's creators that the data does not violate privacy of riders and their trips~\cite{gbfs_specs}.
\item Third, \textit{we recommend changes to the MDS and GBFS specifications by proposing an algorithm
for publishing scooter data with formal privacy and utility guarantees}, which could allow city authorities to
meet their use cases while preserving riders' privacy (\autoref{sec:geo}).
We provide a preliminary analysis on the trade-offs between privacy guarantees and possible utility goals of city authorities.
\end{itemize}

\section{Background: MDS and GBFS}
	\label{sec:background}

LADOT proposed MDS~\cite{mds_specs} in November 2019 as 
a standardized approach for mobility service providers to give trip and event level
information to the city and for the city to regulate mobility service providers.
The proposed use cases include enforcing device cap limits, determining when
scooters are parked in illegal areas, ensuring equitable deployment of devices
across neighborhoods, and informing the city planners' future capital investments and infrastructure
planning~\cite{mds_specs}.

This proposal contrasts with an existing General Bikeshare
Feed Specification (GBFS)~\cite{mds_specs}, which has already been implemented by most mobility companies~\cite{gbfs_systems}.
GBFS specification is designed to provide the status of a
micro-mobility system at the moment (e.g., the number and locations of scooters
that are currently unused). Its
goal is to meet ``data needs for oversight and planning'' while
``protecting traveler privacy'' ~\cite{gbfs_specs}.

The specification for GBFS~\cite{gbfs_specs} requires service providers
to publish certain data feeds as JSON files. Out of the 4 data feeds that are
required, the one that is relevant for our privacy discussion is \emph{free\_bike\_status.json},
which provides a near real-time information on the absolute location of all currently parked scooters.
The feed is updated every TTL seconds. Although no specific value
is recommended in the specification, 60 or 300 seconds is used as the TTL in the datasets we analyze.

MDS, in addition to requiring the implementation of GBFS, also requires
providing a historical view
of all trips and events related to the operation of scooters.
The API specification for MDS has three components, called \emph{Provider},
\emph{Agency}, and \emph{Policy}; each one with a different use case.
Provider API is to be implemented by mobility service providers, and gives a
historical view of all past trips and events on demand. Agency API is implemented by regulatory agencies
to provide end-points that service providers must call to upload events in near real-time
(such as a trip starting or ending), and to provide telemetry data (such as
scooter location, speed, and heading). Policy API is to be implemented by
regulatory agencies to enable them to publish machine-readable policies (such as
localized caps, excluded zones and speed restrictions), and update the policies as needed.
The Provider and Agency APIs provide very granular trip- and event-level
information to the city authorities. The requirements of these two APIs are the ones that have raised privacy concerns in MDS.

\section{Data Collection and Ethics}

For our experiments, we collected data from four scooter companies (Spin, Bolt, Lyft
and Bird) that publish GBFS data feeds for their scooters in Los Angeles. For each company,
we identify the URL for its \emph{free\_bike\_status.json} data feed, and scrape this URL
every TTL seconds. Spin, Bolt, and Bird use a TTL value of 60 seconds, while Lyft uses
300 seconds. All results presented are based on data we collected over 10
days between 2019-11-09 and 2019-11-18.

Our goal is to show that there are privacy risks
in implementing the GBFS and MDS APIs as specified, and propose a solution
that works for both the scooter companies and the regulatory agencies.
Thus, we limit our work to the minimum data collection and analysis necessary
to show that the privacy risks for data published under GBFS are not merely theoretical
possibilities but tangible risks. Consequently, we refrain from associating trips to specific individuals except for trips we took ourselves.

\section{Using GBFS to meet use cases}
\label{sec:you_have_what_you_need}

Here we show that GBFS already provides information
needed to meet the goals of LA's two main use cases: determining the number of scooters
in operation and their distribution across neighborhoods.
We focus on these two use cases out of the 11 listed by~\cite{mds_specs}
because they form the basis for many of the other use cases such as enforcing device caps,
guiding future capital investments based on distribution of scooter usage and informing policy making.
Studying other use cases such as enforcing parking regulations is left as
an area of future work\footnote{It would be interesting to examine whether the precision
provided by GPS coordinates is sufficient for use cases such as enforcing parking rules, regardless of privacy.}.

\subsection{Use Case 1: Determining the Number of Scooters}

The first use case for which the City of Los Angeles requests MDS data is
to determine how many scooters are operating~\cite{mds_specs}.
This helps the city enforce device cap limits per company for a given
area. We show that the GBFS specification already includes data
that can be used to closely estimate the total number of scooters in operation.

\begin{figure}
	\includegraphics[width=1\columnwidth]{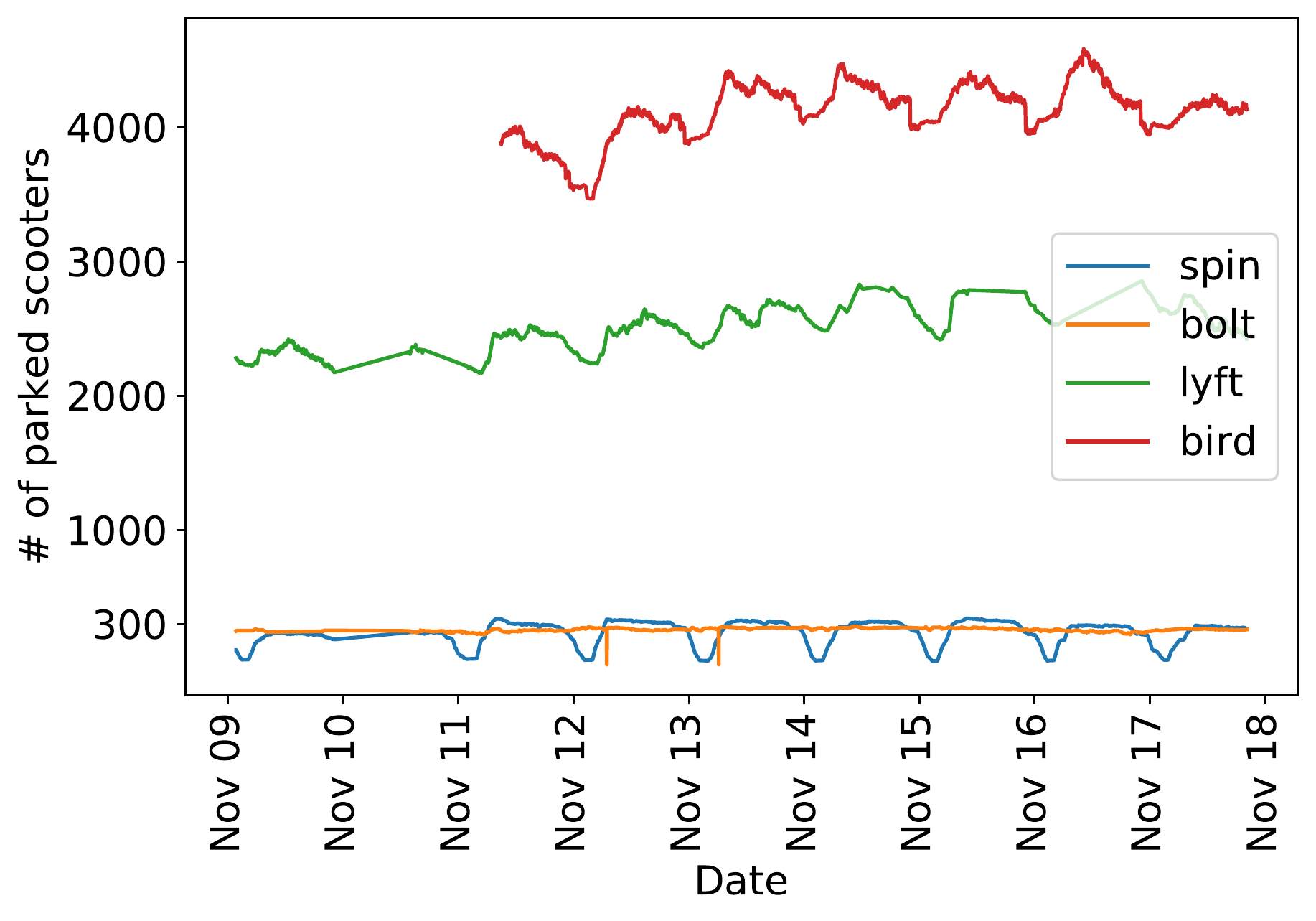}
	\caption{Number of parked scooters for different companies during 2019-11-09 to 2019-11-18}
	\label{fig:num_of_parked_scooters}
\end{figure}

\autoref{fig:num_of_parked_scooters} shows a time series of
the number of parked scooters for four companies: Bird, Lyft, Bolt, and Spin.
The peaks in these plots, which represent times of inactivity, can be
used to estimate the total number of scooters in operation.

LADOT permits each company to run 3,000 scooters, with an additional 2,500 allowed
in disadvantaged communities~\cite{ladot_caps}.
Based on~\autoref{fig:num_of_parked_scooters}, Spin, Lyft, and Bolt are well below
the 3,000 limit. For Bird, one needs to look at the distribution of scooters to see whether the
scooters beyond the 3,000 count are deployed in disadvantaged communities (the next use case).

\subsection{Use Case 2: Determining Scooter Distribution}
The other use case for MDS data that we analyze is the distribution of scooters
within LA neighborhoods. The goal of the city could be to verify that
scooters are operating only in allowed regions or to verify scooters are being
deployed equitably across neighborhoods. The city of LA incentivizes scooter
companies to not only target tourist areas but also to provide their services to less
privileged city areas.

We claim the City of Los Angeles already has enough data available in GBFS to
meet this use case. We demonstrate this by counting the number of parked scooters
in LA neighborhoods for a single GBFS data snapshot. To determine neighborhood limits, we use
shapefiles provided by the City~\cite{LA_shapefiles}. 

\begin{figure}
	\includegraphics[width=1\columnwidth]{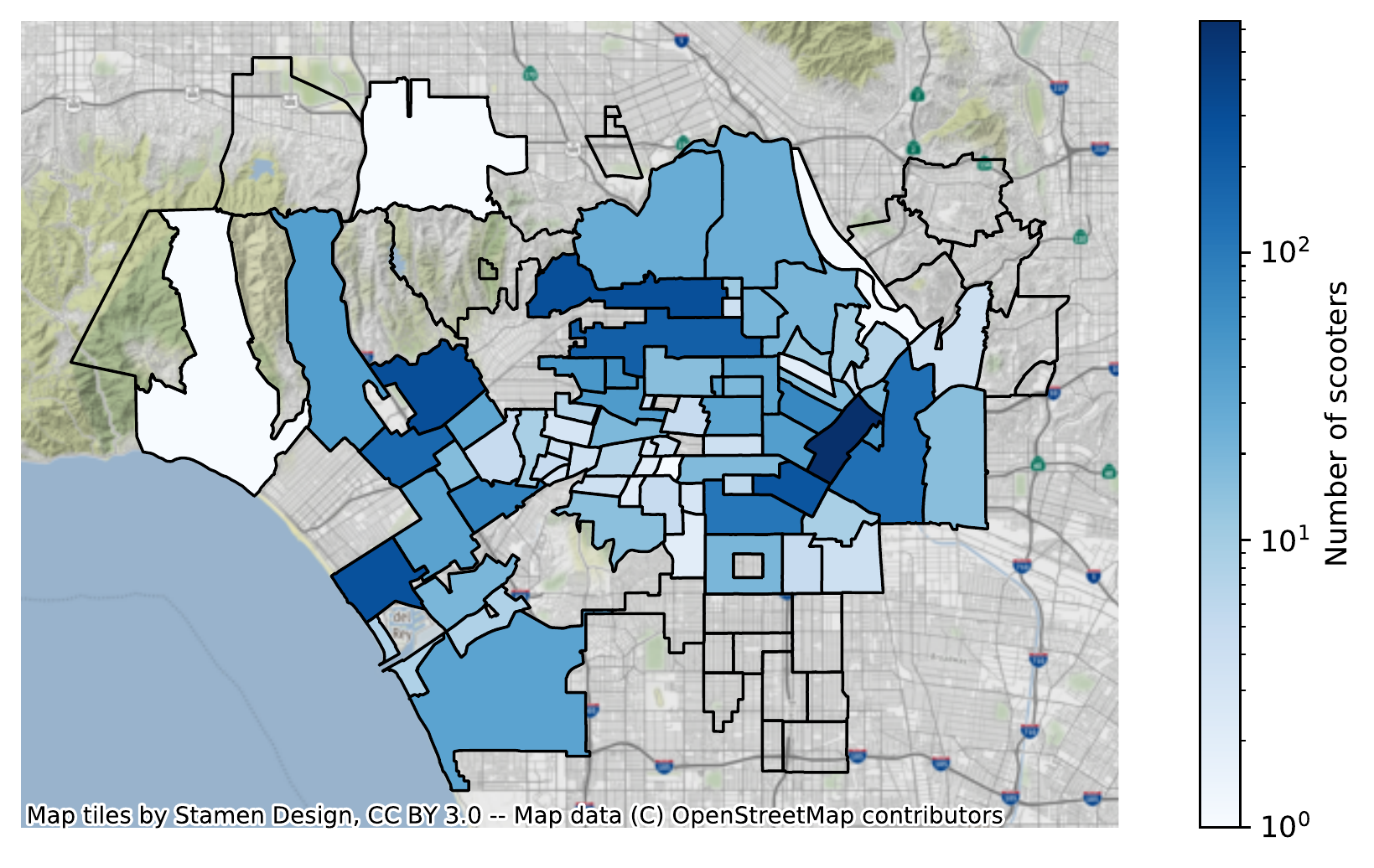}
	\caption{Bird scooter distribution, Los Angeles, 2019-11-12 2:54am}
	\label{fig:bird_scooter_distribution}
\end{figure}

In \autoref{fig:bird_scooter_distribution} we show Bird's scooter distribution
on 2019-11-12 at 2:54am. Bird was the first scooter provider in LA and the one with the
largest number of available scooters in the city. We see three major areas where
Bird favors deploying scooters: Santa Monica - UCLA campus, LA Downtown - USC
campus, and Hollywood. That is, scooters are more prevalent in areas frequented by tourists and students,
and less common in poorer areas.  Therefore, this figure shows that GBFS gives
exactly the information
needed for the use case of determining scooter distribution across neighborhoods.

\section{Attacks on GBFS}
	\label{sec:attacks}
We next show how the access to data in GBFS violates the privacy of scooter users.
We demonstrate this by reconstructing trips using parked scooter data and
clustering the trips to identify interesting trips, which we then further investigate.

\subsection{Reconstructing Trips}
	\label{sec:recon}
	
We use parked scooter data in GBFS to reconstruct trips,
showing that GBFS, despite publishing
much less information about scooters than MDS, does not protect
the privacy of scooter users.

We check the location of all parked scooters
on a minute by minute basis. If the location of a scooter changes, we record the scooter's last location
and the last time it was this location as the start location and start
time of a trip, respectively. We record the scooter's new location and the
first time it was this location as the end location and end time of the trip, respectively.

This process provides us a dataset of trips, including potentially ``fake trips'' that may be
due to a scooter being relocated for maintenance or charging purposes.
After looking at the overall distribution of trip distances and durations,
we filtered out such potentially fake trips using a trip distance threshold of 100 meters minimum
(to filter out slight relocations of a scooter) and a time duration threshold of
1 hour maximum (to filter out relocations due to charging or maintenance)\footnote{More sophisticated techniques can be used to filter out fake trips
but they are not necessary for our proof-of-concept demonstration.}.

We looked for specific trips as a means to validate our trip reconstruction.
To accomplish this without violating other people's privacy, we took
scooter rides ourselves, and recorded the time and location for both
start and end points.
We were able to match the time and location of trips we took with
a specific record in the trip dataset, which we reconstructed using the method described above.
We are able to do this simply because we have information regarding the time and location our trips start and end at.
This approach, in addition to validating our reconstruction, demonstrates the risk of identifying trips by other individuals
when side information is available. Studying how different sources of side information,
which prior works have shown are often easily available ~\cite{nyctaxicabs, nytimes_apps},
can be used to match trips with individuals is an interesting area for future work.

\subsection{Identifying Visits to Interesting Destinations}

Certain habits and preferences, although legal, are considered personal and private.
 GBFS data allows any member of the public access to data related to these habits.
We show how we use clustering on the reconstructed trips to identify
popular destinations and learn about trips to specific places.

\subsubsection{Detecting Interesting Spots by Clustering}
\label{sec:clusters}

We use clustering to identify hotspots.
The top clusters of parked scooters indicate popular places to start and end a scooter ride.
Such statistics may not contain much private information if the hotspot is an area such as 
a popular shopping destination, which would obviously bring many scooters. Tracking an individual 
scooter may be hard in these areas. However, if the hotspots are near houses or other
establishments, they may help us infer the habits of individual users.

A basic attack begins by associating the start and end locations of trips. Trips may end 
in a parking hotspot in a public area or at a specific house or establishment. By finding
a cluster of trips ending at a person's home, we can check the start and end locations of 
those trips to find out where the person is going to or coming from. Moreover, this data includes
timestamps, thus we can infer more information about these trips.
End points can be more sensitive than start points, as people might need to walk a certain distance to 
find a scooter before starting a trip, but they typically end a trip at their
exact destination.

\begin{figure}
\includegraphics[width=0.5\textwidth]{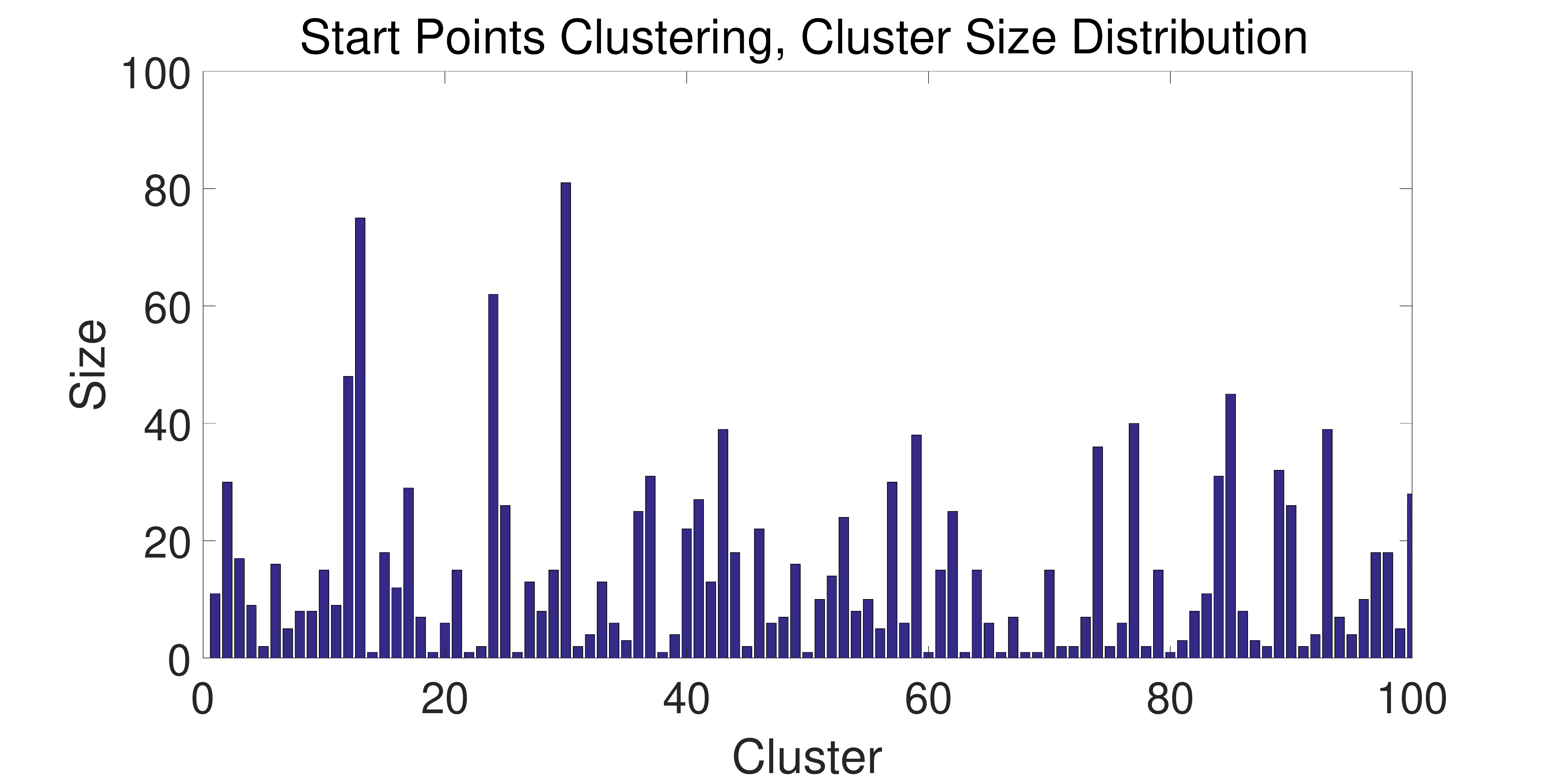}
\caption{Distribution of 100 clusters on start locations, Spin}
\label{fig:clustering_100_start_distribution}
\end{figure}

We use the standard k-means clustering algorithm on the start locations. 
By increasing the target number of clusters, we can make each cluster quite small and 
concentrate it in a narrow geographic area, which we can use to identify particular 
places. \autoref{fig:clustering_100_start_distribution} shows the distribution of cluster
sizes when we have 100 clusters. Since large clusters may be popular public parking areas, 
we concentrate on smaller clusters with about 10 trips, which may contain more interesting information.
In the 10 days of collected data from a company with a small number of scooters, such as Spin,
it is reasonable to assume that a person takes a ride every day. 
Cluster 54, with size 8, centering in [34.02019, -118.27633], is located directly next to
a marijuana dispensary. Below, we will explore trips to this location.

\subsubsection{Case Study: A Cluster Outside a Marijuana Dispensary}
\label{sec:attack}

Based on our clustering (see section \ref{sec:clusters}), we identified a
marijuana dispensary near a university campus, located within a 2km distance of the university campus and
student housing areas. The lack of other stores or apartments on the same block as the dispensary,
helps to identify the marijuana store as the likely ride destination of nearby parked scooters.

\begin{figure}
	\includegraphics[width=1\columnwidth]{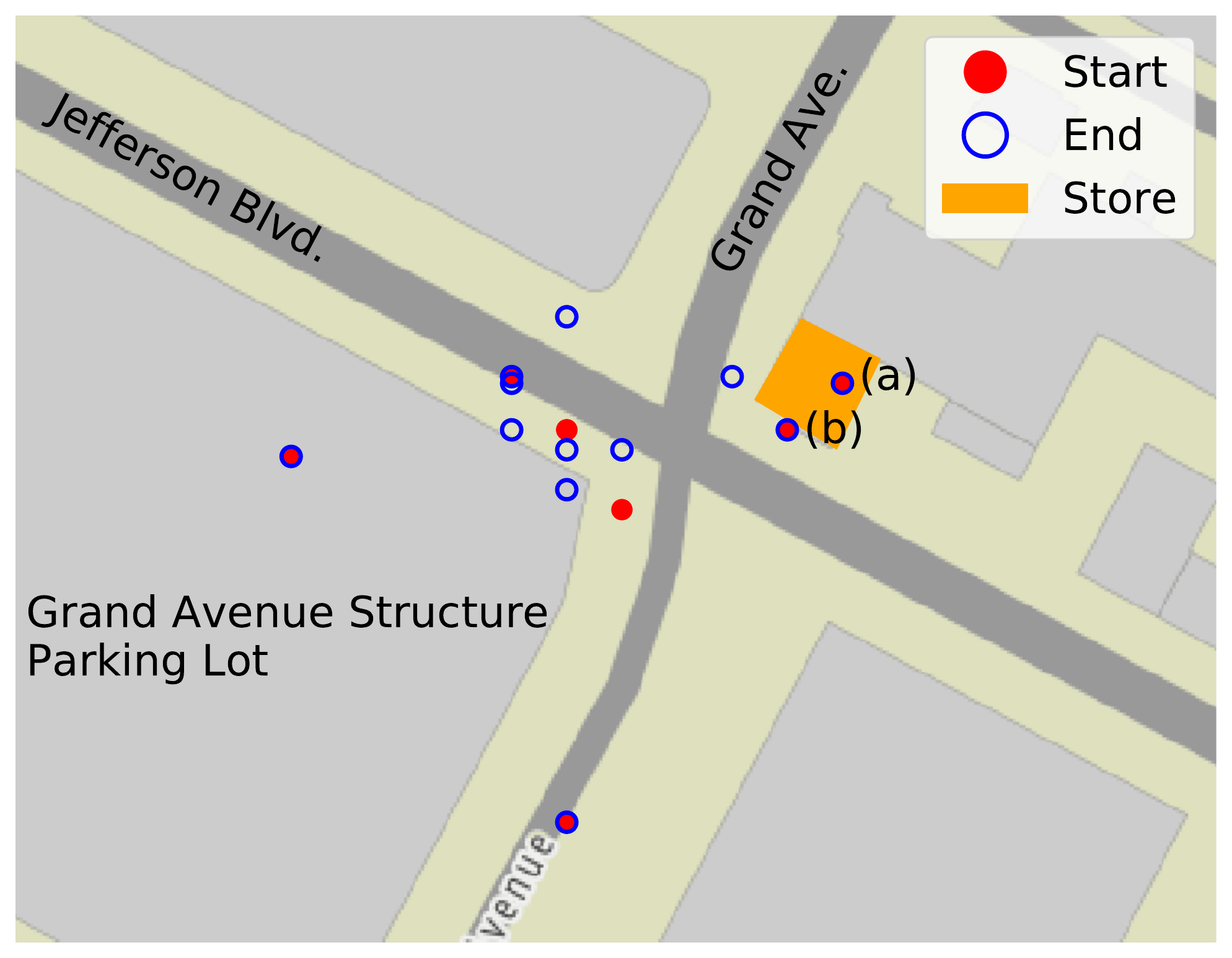}
	\caption{Spin trips starting and ending in Jefferson Blvd. and Grand
Ave. between 2019-11-11 and 2019-11-12}
	\label{fig:spin_locations}
\end{figure}

In \autoref{fig:spin_locations} we show Spin trips starting and ending close to
the dispensary between 2019-11-11 and -12. We focused on two locations, i.e., (a) and
(b), which are the end of a trip followed by the start of a second trip a few
minutes later. More specifically, for (a) the trip ends at 2019-11-11 11:32am, and
resumes 5 minutes later. For (b) the trip ends at 2019-11-12 3:08pm and resumes
44 minutes later. In case (a), the time difference between the end of the
first trip and the start of the second trip provides sufficient time to make a
purchase, but is short enough to suggest the same rider.

For ethical reasons, we do not check where the trip originates before a rider reaching the store,
and where it ends after a scooter user leaving the store, although doing so is possible using the data.

Moreover, going beyond checking the starting location of a trip,
one can potentially correlate the trip start and end points with specific riders using auxiliary data such as public records for inhabitants of particular addresses.

One could also gain more confidence in the inferred trips and their correlation to specific individuals through a longitudinal study
focusing on recurrent trips. In other words, a malicious adversary could use the GBFS data
to learn about personal habits of particular individuals using the scooters.

\subsection{Attacks without using scooter IDs}
\label{sec:more_attacks}

The trip reconstruction described in section~\ref{sec:recon} assumes a scooter's
ID does not change over time, which is true for the Spin dataset we used.
Some companies, such as Bird, change their scooter IDs frequently, despite the GBFS specifications
requiring IDs to be consistent over time. We explain why the changes in scooter IDs may not be sufficient to prevent privacy violations. 
We give a high-level description but leave exploration of possible attacks as an area the future work.

The simplest method of attack is to find a scooter that is isolated from other scooters either in time (it is the only scooter that moved)
or distance (it is in an isolated area). We hypothesize this scenario can be generalized
to the case of multiple scooters. By using the general
distribution of duration and distance of scooter trips, we can build a probabilistic model and compute possible 
matchings to track scooters over time. The ground truth can be obtained from some datasets whose IDs do not change. For example, we can train the 
model on Spin data and then use it to predict trips on Bird data.

In summary, if locations of scooters are not reported in a privacy-preserving manner, even without using stable IDs,
this type of attacks may succeed. That is why, in the next section, we recommend using techniques with formal privacy
guarantees to report noisy, rather than precise locations, which would increase the difficulty of privacy-violating inferences.

\section{Geo-Indistinguishability as a potential solution}
	\label{sec:geo}

In this section we discuss one way to define location privacy and a mechanism through which we can achieve this type of privacy. We apply this mechanism 
a restricted set of MDS's use cases and evaluate its performance. In addition, we provide suggestions for improving GBFS and MDS.

\subsection{Definition and Mechanism}
A basic technique to protect location privacy is to report a noisy location. Intuitively, if two locations are close to each other, it can be difficult to 
distinguish between them during the observation.
Geo-Indistinguishability (G-I)~\cite{andres2013geo} is a popular location privacy definition that provides protection to such location pairs.
Formally, for any two locations $x$ and $x'$ with a maximum distance $R$, the 
observation $S$ of them is $(\epsilon,R)$ indistinguishable if the following condition is satisfied:
\begin{equation*}
\frac{\Pr(S|x)}{\Pr(S|x')}\leq e^{\epsilon R}, \forall R>0,\forall S, \text{ } \forall x,x':d(x,x')\leq R
\end{equation*}
The definition is parameterized by the radius $R$ and the privacy loss parameter $\epsilon$. Note that 
$\epsilon$ contains the unit of measurement, while the $\epsilon$ in the traditional differential 
privacy definition\cite{dwork2006calibrating} does not. In this work, we choose kilometer as the unit of measurement. For example, if 
we choose $R=0.2,\epsilon=\frac{\ln 4}{R} = 5~{\ln 4}$, that means for any two locations $x$ and $x'$ which 
are within 0.2km, the probabilities of them having the same observation $S$, $\Pr(S|x)$ and $\Pr(S|x')$, the ratio of the two probabilities is upper bounded by 4. Namely, $\Pr(S|x)\leq 4\Pr(S|x')$ and $\Pr(S|x')\leq 4\Pr(S|x)$.

Algorithm \autoref{alg:geo} shows the basic framework of a mechanism that achieves G-I. For implementation details, see Appendix \ref{sec:geoappendix}.

\begin{algorithm}
	\label{alg:geo}
	\SetKwInOut{Input}{Input}
	\SetKwInOut{Output}{Output}
	
	\Input{$\epsilon$, $R$, location $(lat, lon)$}
	\Output{Noisy location $(lat', lon')$}
	Sample angle $\theta$ and length $r$ according to $\epsilon$, $R$.
	
	Move the location $(lat, lon)$ by $(\theta,r)$.
	
	Output $(lat', lon')$.
	
	\caption{Geo-Indistinguishability~\cite{andres2013geo}}
\end{algorithm}

\subsection{Utility Analysis}
In this section we define utility and show that for reasonable privacy
parameters, a city can allow scooter companies to use G-I and still satisfy its goals
with only a moderate increase in
measurement error.  We consider use cases discussed
in section~\ref{sec:you_have_what_you_need}: (1) total number of scooters within
city limits, and (2) scooter distribution by neighborhood, and evaluate these use cases 
using publicly available scooter location data with G-I.

We measure utility as the difference between the true number of scooters within
a bounded region and the perturbed number of scooters within that region due to noisy location reporting.
A smaller difference reflects a better utility.
Our expectation is to see a utility drop as privacy guarantees are increased.
For example, more congested neighborhoods losing scooters to lesser populated
ones (use case 2), or scooters leaving the overall city limits (use case 1).

\begin{figure*}
	\includegraphics[width=1\columnwidth]{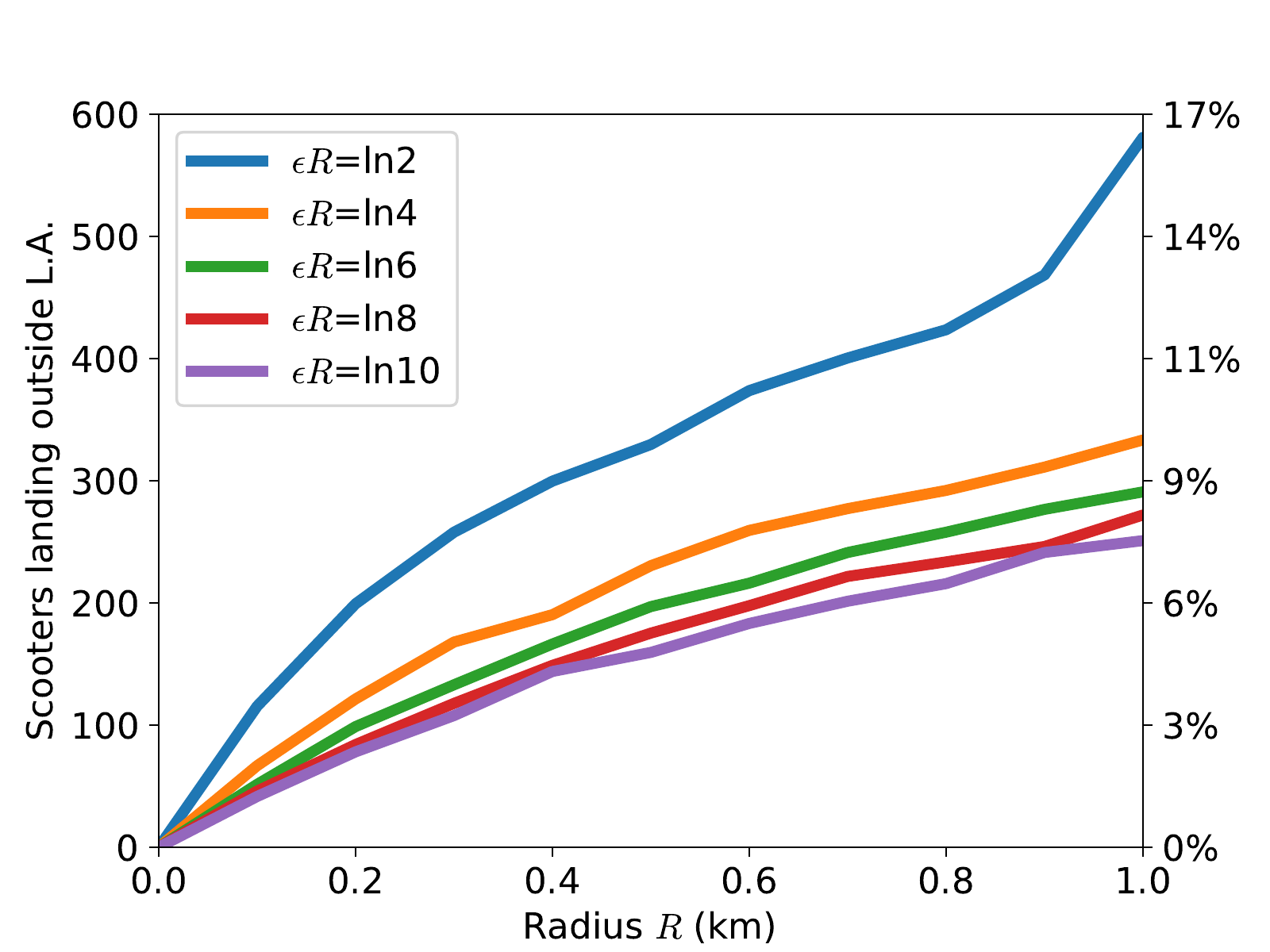}
	\includegraphics[width=1\columnwidth]{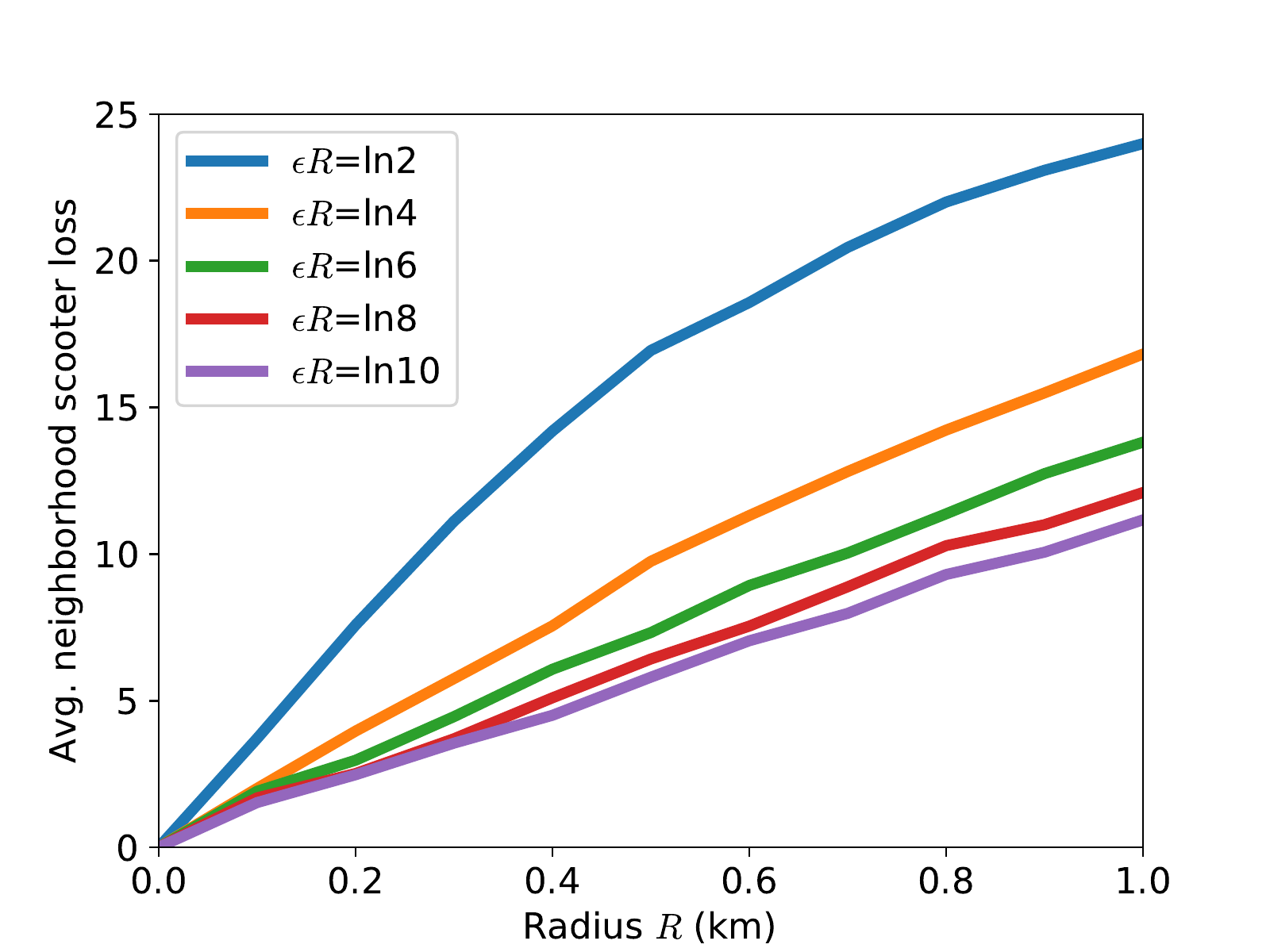}
	\caption{Number of scooters landing outside LA  boundaries (left) and
average scooter loss in LA neighborhoods (right) when adding
noise. Based on Bird Parked scooters 2019-11-12 2:54am}
	\label{fig:privacy_vs_utility_total_bird}
\end{figure*}

To evaluate the impact of G-I on use case 1, we perturb parked scooter
locations for different values of radius $R$, varying it from 0 to 1 km in increments of 0.05 km.
The repeat the procedure 100 times for each value of $R$, and record the average
number of scooters leaving LA boundaries.
\autoref{fig:privacy_vs_utility_total_bird} (left) shows the number of Bird scooters
outside city boundaries using a data snapshot corresponding to 2019-11-12 02:54am.
These scooters randomly moved to neighboring cities, or even into the
ocean. The figure shows that, as expected, as the privacy guarantees (and hence the noise) increase, the number of scooters incorrectly placed increases in an upward
linear trend.

To evaluate the impact of G-I on use case 2, we consider the average
number of scooters lost per neighborhood; in this case 108 neighborhoods. As 
\autoref{fig:privacy_vs_utility_total_bird} (right) shows,
neighborhood loss follows
a similar upward trend as in use case 2, where adding more noise results in
higher error.

In the context of densely populated cities such as LA, a radius $R$ that is
reasonably large and covers multiple blocks can help obfuscate specific destinations.
We pick 0.25 km as the radius for our privacy protection, which is comparable to two and a half blocks, and
analyze how the algorithm performs with this parameter. 
With $\epsilon=\frac{{\ln 6}}{R} = 4~{\ln 6}$, 
in \autoref{fig:privacy_vs_utility_total_bird}
we see a total loss of 115 scooters (3\%) (left), and an average of 4 scooters
lost per neighborhood (right). This suggests that for reasonable privacy
parameters the utility loss is not catastrophic.

\begin{figure}
\includegraphics[width=0.5\textwidth]{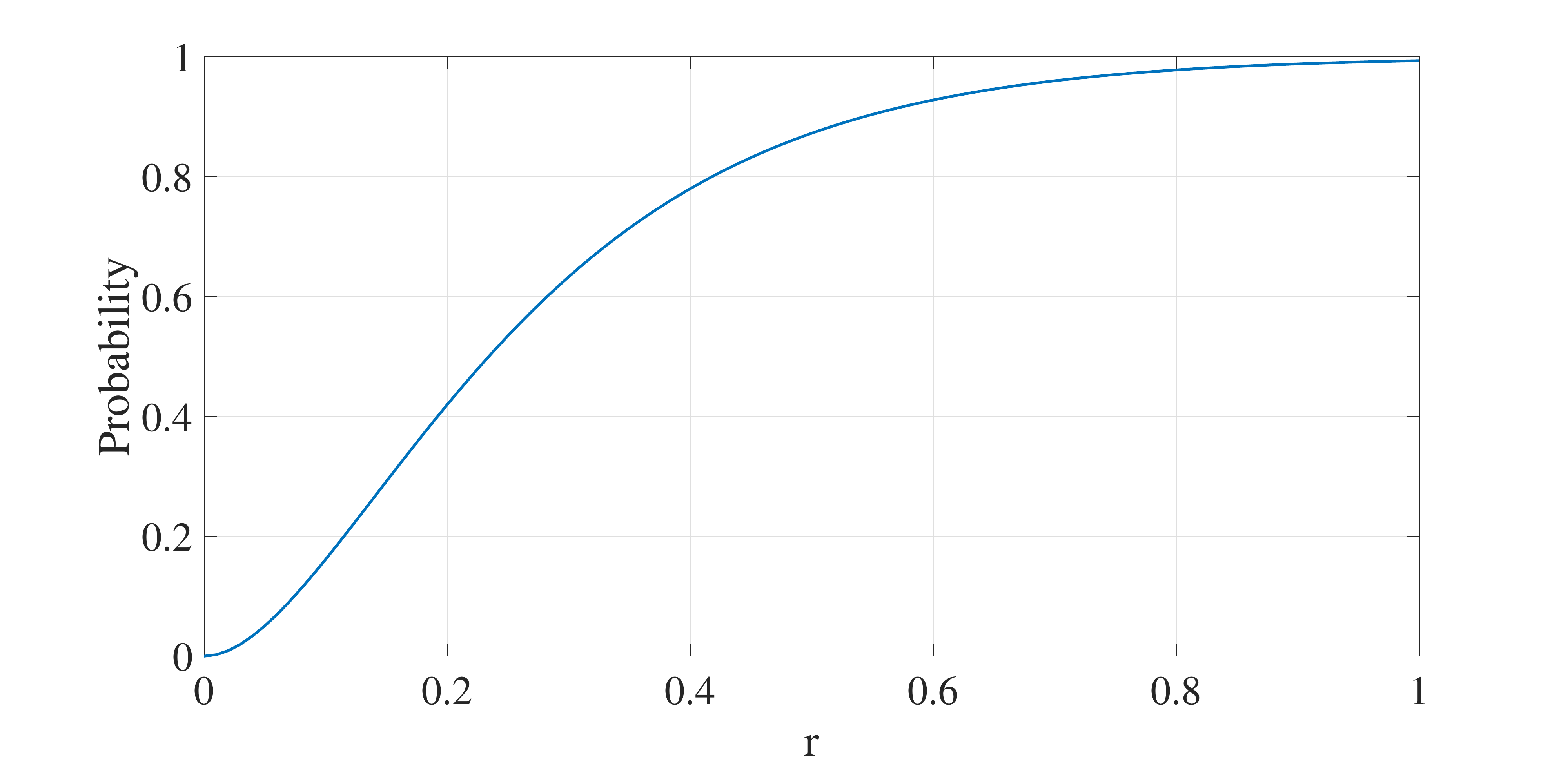}
\caption{Cumulative function for $R=0.25, \epsilon= 4\ln 6$}
\label{fig:cdf}
\end{figure}

In addition to the above empirical analysis, we theoretically calculate
the utility provided by the values we chose for $R$ and $\epsilon$.
We plot in~\autoref{fig:cdf}
the probability of a sampled noisy location being within a certain distance
from the true location. The cumulative function $F_\epsilon$ is given by the following
equation~\cite{andres2013geo}: 
\begin{equation*}
F_\epsilon(x)=1-(1+\epsilon x)e^{-\epsilon x}
\end{equation*}
As the figure shows, about 50\% of the noisy locations will land within 250m
of the true locations, while 99\% of the noisy locations will land within a 1 km
distance. This implies that in expectation, the parameters we picked in the mechanism do not greatly change the true locations.

\subsection{How should GBFS and MDS change?}
	\label{sec:recs}

Having demonstrated an approach towards publishing GBFS data with privacy guarantees and reasonable utility loss,
we recommend the debate between scooter companies and the cities be resolved by using rigorous state-of-art techniques for privacy-preserving data publishing.
GBFS data feeds could use this mechanism to report the noisy locations of scooters, and
MDS requirements could be updated to remove the trip and event level data requested in
Provider and Agency APIs. The Policy API does not require the service providers to publish
any data, so we do not recommend any changes to the Policy API.  
Both sides of the debate, and, most importantly, the riders, would benefit 

if scooter companies would be willing to invest
resources in implementing privacy-preserving mechanisms, and cities would be open to 
considering performing their analyses on privacy-protected data.

\section{Related Work}
Privacy challenges of sharing location data have been well-documented in the literature. 
For instance, Culnane \emph{et al.}~\cite{culnane2019stop} show the vulnerabilities of
Melbourne's public transportation by identifying co-travelers and complete
strangers when considering the uniqueness of each record within the dataset.
Other datasets, such as the New York City taxicab dataset~\cite{nyctaxicabs} and
geolocation data gathered by apps from mobile devices~\cite{nytimes_apps}, have been shown to have similar privacy risks. 
In addition, Bassolas \emph{et al.}~\cite{bassolas2019hierarchical} show that sharing precise user locations may carry privacy risks.

In the last two decades, researchers have demonstrated approaches for mitigating the privacy risks of location data sharing\cite{bassolas2019hierarchical,primault2018long,andres2013geo}.
For example, the recent work of Bassolas \emph{et al.}~\cite{bassolas2019hierarchical} proposed a framework for 
analyzing the hierarchical mobility structure of cities. In order to greatly reduce privacy risks, their data of trip flows are anonymized and aggregated. Primault \emph{et al.}
\cite{primault2018long} provide a detailed survey of the existing location privacy 
protection mechanisms. 

In this work, we based our proof-of-concept solution approach on Geo-Indistinguishability~\cite{andres2013geo},
which is a definition that aims to relax the notion of differential privacy for location-specific use cases.

\section{Conclusion}
	\label{sec:conc}
	
Our work demonstrates that the data fight between cities and mobility data providers could benefit from adoption of state-of-the-art 
privacy-preserving techniques. The adoption of such techniques would require flexibility both from the side of scooter companies and 
the cities. However, the result would be increased privacy for the riders while providing cities the ability to achieve their regulatory goals.

\section{Acknowledgements}
This work was supported, in part, by NSF Grant \#1755992.

Guillermo Baltra's research is sponsored in part by the Department of
Homeland Security (DHS) Science and Technology Directorate, Cyber
Security Division (DHS S\&T/CSD) via contract number 
70RSAT18CB0000014,
and by Air Force Research Laboratory under
agreement number FA8750-18-2-0280.
The U.S.~Government is authorized to reproduce and distribute
reprints for Governmental purposes notwithstanding any copyright
notation thereon.

\bibliographystyle{IEEEtran}
\bibliography{references}

\appendix
\subsection{Geo-indistinguishability}
	\label{sec:geoappendix}
In this section we talk about the details of generating the noisy location. 
The privacy preserving algorithm \autoref{alg:geo}{ randomly selects the angle $\theta$ uniformly. Then length $r$ is drawn from Laplace distribution. 
$\theta$ tells which direction we should go and $r$ represents the distance in that direction.
With the pair $(\theta,r)$, we move from the original location and obtain a noisy location. In total, this pair is sampled from zero-centered polar 
Laplacian.
\begin{equation*}
D_{\epsilon}(r,\theta)=\frac{\epsilon^2}{2\pi}r e^{-\epsilon r}
\end{equation*}
According to the analysis in \cite{andres2013geo}, algorithm \autoref{alg:geo} using the above sampling technique preserves $(\epsilon, R)$ Geo-Indistinguishability.

We sample $r$ from the above distribution 
and sample $\theta$ uniformly from $[0, 2\pi]$.
Since we are working on geo-coordinates, we need to use the following formula to compute the noisy location
(assuming $R_{earth}=6378.1km$). Given latitude $x_{lat}$ and longitude $x_{lon}$ for a location, the noisy 
latitude $x_{lat}'$ and longitude $x_{lon}'$ can be computed by:

\begin{equation*}
x_{lat}'=\arcsin(\sin x_{lat} \cos\frac{r}{R_{earth}}+\cos x_{lat} \sin\frac{r}{R_{earth}}\cos\theta)
\end{equation*}
\begin{equation*}
x_{lon}'=x_{lon}+\arctan\frac{\cos\frac{r}{R_{earth}}-\sin x_{lat}\sin x_{lat}'}{\sin\theta\sin\frac{r}{R_{earth}}\cos x_{lat}}
\end{equation*}

\end{document}